# Two-Qubit Geometric Phase Gate for Quantum Dot Spins using Cavity Polariton Resonance

Shruti Puri[1], Na Young Kim[1] and Yoshihisa Yamamoto[1,2]
1. E. L. Ginzton Laboratory, Stanford University, Stanford, California 94305, USA and
2. National Institute of Informatics, 2-1-2 Hitotsubashi, Chiyoda-ku, Tokyo 101-8430, Japan
(Dated: March 21, 2012)

We describe a design to implement a two-qubit geometric phase gate, by which a pair of electrons confined in adjacent quantum dots are entangled. The entanglement is a result of the Coulomb exchange interaction between the optically excited exciton-polaritons and the localized spins. This optical coupling, resembling the electron-electron Ruderman-Kittel-Kasuya-Yosida (RKKY) interactions, offers high speed, high fidelity two-qubit gate operation with moderate cavity quality factor $Q$. The errors due to the finite lifetime of the polaritons can be minimized by optimizing the optical pulse parameters (duration and energy). The proposed design, using electrostatic quantum dots, maximizes entanglement and ensures scalability.

The base of a quantum computer is a qubit, which is essentially any physical system that can store a linear superposition of basis states. Quantum computation is impossible without a coherent interaction between two qubits [1]. Electron spin trapped in a quantum dot (QD) is one of the most promising systems from the viewpoint of single qubit operations. The complete ultrafast optical control of the spin of such localized electron has been demonstrated [2]. A two-qubit gate, on the other hand, requiring an interaction between two electron spins is yet to be demonstrated. Coupling mechanisms based on the spatial overlap of the wavefunctions of the electrons, for example, by the gating of the tunneling barrier between neighboring dots have been proposed [3, 4]. However, the major drawback of these schemes is that the interaction is short ranged and the gate operation is slow for a reasonable separation.

A two-qubit geometric phase gate based on the Coulomb exchange interactions between the electrons, mediated by exciton was proposed [5]. This interaction based on itinerant optical excitations is analogous to the RKKY coupling in which an indirect exchange interaction between two magnetic impurities is mediated by itinerant electrons [6]. Exciton-polaritons, being four-five orders of magnitude lighter than bare excitons in mass, have reduced interaction with the environment (for example phonons) [7] and hence are better suited as interaction medium. A similar entanglement scheme based on optically excited microcavity polaritons in a quantum well (QW) has been proposed, in which electron spins separated by hundreds of nanometers are coupled [8]. We propose a new system, in which the electron spins are trapped in further separated electrostatic QDs. The QW is placed in a planar micro-cavity, which is designed to facilitate selective excitation of single transverse-mode polariton and therefore increases the interaction energy by nearly two orders of magnitude, for electron spins separated by distance on the order of a few micrometers.

Geometric phase gates offer the advantage of being resistant to random fluctuations from the environment and thus robust to random noise. Since the geometric phase depends only on the area enclosed by the path along which the system evolves and not the actual trajectory, random fluctuations average out to zero along the loop [9]. Hence, any error arising due to random fluctuations is eliminated. A geometric phase gate is also robust against errors due to the dynamical phase noise [10].

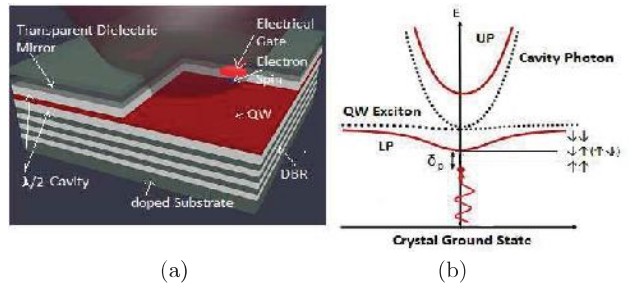

FIG. 1. (a) Illustration of the system consisting of QW placed in $\lambda/2$ DBR cavity. Two electrons are trapped by the electric filed under the metal gates. Polaritons are formed in the region between the dots where the laser is incident. (b) Representation of the exciton polariton energy dispersion. The inset shows the spin dependent energy of the LP.

Creating entanglement between two qubits is not sufficient for a two-qubit gate to be usable in a quantum logic circuit. Firstly, the gate must have high fidelity, i.e., it should be free of errors. In addition, no information about the state of the qubits must be leaked. This paper particularly discusses these important aspects of a two-qubit gate in a real setup, which has significant losses that cause decoherence and leak information about the system. It has been shown that a single qubit rotation and a controlled rotation of two qubits like a controlled-$z$ rotation form a set of universal logic gates [11]. In this paper we have achieved a two-qubit controlled-$z$ gate. The scheme for a single qubit gate based on principles described here will be presented elsewhere.

The proposed system consists of a single GaAs QW, bounded by AlGaAs layers, placed in a $\lambda/2$ cavity with a transparent dielectric mirror on the top and a AlAs/AlGaAs distributed Bragg reflector (DBR) at the bottom. The electric field profile due to positive potential at the transparent metal electrode, placed on the top of the QW, creates an electrostatic QD (as shown in Fig.1(a)). A single electron from the n-doped AlGaAs layer at the bottom is trapped in each QD [12]. The planar microcavity is resonant with the excitonic transition of GaAs QW. As shown in Fig.1(b), in strong coupling regime, the normal states are the quasi-particles: lower polariton (LP) and upper polariton (UP), formed by the admixture of cavity photons and optically active heavy-hole excitons [13]. The splitting between the LP and UP branch depends on the strength of the coupling between the cavity and the exciton. The circularly polarized ($\sigma+$) pump laser is focused in a small spot (radius = $R$) parallel to the growth direction i.e., $\mathbf{k}_{||} = 0$, covering only two QDs. The optically excited excitons consist of electron spin $j_{ze} = -1/2$ and hole spin $j_{zh} = 3/2$. The excitons with $j_{ze} = 1/2$ and $j_{zh} = 3/2$ are called dark excitons since they cannot couple with the incident photons. Our scheme virtually excites only LP with zero transverse momentum, so the frequency of the laser, $\omega_L$ is off-resonant (red detuned) from LP energy at $\mathbf{k}_{||} = 0$.

The localized electron spin interacts with the electronic part of the polariton via Coulomb interaction. The direct part of the interaction just leads to a renormalization in energy. The exchange part is spin dependent, which results in coupling between the two spatially separated electron spins mediated by exciton-polaritons. The Coulomb exchange interaction with the hole is very small and hence neglected [8, 14]. The Hamiltonian for the interaction between the localized electron and the polariton in the rotating wave approximation is [8]:

$$H_0 = \delta_x b^\dagger b + \delta_p p^\dagger p + V\{(s_{1z}+s_{2z})b^\dagger b - r_0^2(s_{1z}+s_{2z})p^\dagger p \\ + r_0(s_{1+}+s_{2+})b^\dagger p + r_0(s_{1-}+s_{2-})p^\dagger b\} + \Omega(t)(p^\dagger + p), \quad (1)$$

where $\delta_x(\delta_p)$ is the dark exciton(polariton) detuning from the laser, $b^\dagger(p^\dagger)$ are the dark exciton(polariton) creation operators, $s_{1z}(s_{2z})$ are the projection of the spin operators in the $z$ direction for the localized electron spins labeled 1(2), $s_{1+(-)}(s_{2+(-)})$ are the spin raising(lowering) operators, $r_0$ is the Hopfield factor for the polaritons ($r_0 = 1/\sqrt{2}$ when the cavity is resonant with QW exciton energy) and $\Omega(t)$ is the external pumping rate. $V$ is the exchange coupling constant between the localized electron spins and the electronic part of the polariton, given as:

$$V = \int d\mathbf{r_e} d\mathbf{r_h} d\mathbf{r_l} \frac{\psi(\mathbf{r_e},\mathbf{r_h})\phi(\mathbf{r_l})e^2\psi(\mathbf{r_l},\mathbf{r_h})\phi(\mathbf{r_e})}{4\pi\epsilon(|\mathbf{r_e}-\mathbf{r_l}|)}. \quad (2)$$

where $\epsilon$ is the dielectric constant of GaAs, $\mathbf{r_e},\mathbf{r_h}$ represent the position vectors of the electron and hole in the excitonic part of the polariton, $\mathbf{r_l}$ represents that of the localized electron, $\psi, \phi$ represent the wavefunctions of the excitonic component of the polariton and localized electron. The origin of the dark exciton terms is due to the scattering process involving a spin flip between the localized electron 1(2) and the electronic part in the polariton. In our system, since the dark exciton resonance is far detuned from the pump laser and the coupling constant $V$ is small, we can safely eliminate scattering into these modes. Thus the Hamiltonian reduces to:

$$H_0 = \delta_p p^\dagger p - Vr_0^2(s_{1z}+s_{2z})p^\dagger p + \Omega(t)(p^\dagger + p) \\ = \delta_s p^\dagger p + \Omega(t)(p^\dagger + p). \quad (3)$$

This implies that the frequency shift of the polariton resonance ($\delta_s = \delta_p - V(s_{1z}+s_{2z})r_0^2, s = s_{1z}s_{2z} \equiv \uparrow\uparrow, \downarrow\downarrow, \uparrow\downarrow$ or $\downarrow\uparrow$) depends on the spins of the two localized electrons. Thus if both spins have $s_z = +1/2$, then the LP resonance will be $\delta_{\uparrow\uparrow} = \delta_p - Vr_0^2$. Similarly if both spins have $s_z = -1/2$, then $\delta_{\downarrow\downarrow} = \delta_p + Vr_0^2$ and if one spin has $s_z = -1/2$ while the other in $s_z = +1/2$, then the detuning will remain the same i.e., $\delta_{\uparrow\downarrow} = \delta_{\downarrow\uparrow} = \delta_p$. This energy level splitting is shown in the inset of Fig.1(b).

It is important to note that in general the inelastically scattered polariton by the localized spins can acquire any $\mathbf{k}$. However, the virtually excited LPs interact with the localized electrons for a short time, so that, to the first order we can neglect the inelastic scattering and only consider elastic scattering. The mediating polaritons thus have zero transverse momentum, leading to a large in-phase interaction between the spatially separated localized electrons. In our analysis we have neglected the real excitation from the trapped electron state to the charged exciton(trion) state due to large detuning. The charged exciton formation in the presence of electric field is suppressed because of small oscillator strength caused by the spatial separation between the hole and electron [15, 16].

If we assume that the electrons are trapped in a parabolic potential of radius $a$ and the exciton-polariton is optically excited into a spot size of $A = \pi R^2$, the two wavefunctions are given by:

$$\psi = \frac{1}{\sqrt{A}}\sqrt{\frac{2}{\pi}}\frac{1}{a_B}e^{-|\mathbf{r_e}-\mathbf{r_h}|/a_B}, \quad \phi = \frac{1}{\sqrt{2\pi a^2}}e^{-|\mathbf{r}-\mathbf{R_i}|^2/2a^2},$$

where $a_B$ is the exciton Bohr radius. Using these wavefunctions in Eq. (2), the interaction energy is given by $V = a/4\pi\epsilon R^2$ (see supplementary information for detailed derivation). It should be noted that if the overlap between the polariton and localized electron wavefunctions is higher, the interaction energy would be greater. Hence, if $R$ decreases (or $a$ increases) $V$ increases.

It is known that virtually excited polaritons stay inside a cavity only during the pulse width but the polaritons decay rapidly from a microcavity due to short polariton lifetime [17]. The photon lifetime in a DBR cavity of effective length $L_c$, refractive index $n_c$ and DBR mirror reflectivity $r$ is $\tau_{photon} = (n_c L_c)/(c(1-r))$, where $c$ is the speed of light. The effective cavity length $L_c$ is a few wavelengths longer than the cavity length $\lambda/2$, due to the penetration of the cavity field into the DBR ($L_c = \lambda/2 + L_{DBR} \approx 5\lambda/2$) [18]. The lifetime of exciton-polaritons is $\approx \tau_{photon}/t_0^2$, with $t_0 = 1/\sqrt{2}$ as the photonic Hopfield factor. The exciton-polaritons have the unique spot size given by [17, 19] $R = \sqrt{(\lambda L_c)/(\pi(1-r)n_c)}$. Note that, in order of have a higher exciton-polariton lifetime (obtained by increasing $r$), the spot size must be increased. Thus, the exciton-polariton lifetime in the cavity and interaction energy are inversely related. For typical reflectivities $r \approx 99.2\%$ corresponding to $Q \approx 500$, $n_c = 3.5$ and $\lambda = 786$ nm, the unique spot size is $R = 2$ $\mu$m and $\tau_{photon} = 0.65$ ps. Thus $\tau_{polariton} = 1.3$ ps or the decay rate $\gamma = 3$ meV. In a single GaAs QW of thickness $\approx$ 8 nm and Bohr radius $a_B = 12$ nm, the exchange coupling energy between two electrons localized in a circular QD of 100 nm diameter, separated by 2 $\mu$m and virtually excited exciton-polaritons in circular region of cross-sectional area $A = \pi R^2 = 12.6$ $\mu$m$^2$ is $V = 2$ $\mu$eV. This value is 100 times higher than that reported in [8] for a separation of about 2 $\mu$m between the electrons. The increase in the coupling strength is due to the absence of higher momentum modes that caused destructive interference by accumulating a phase of $(\mathbf{k} \cdot \mathbf{r_{12}}) = \pi$ while moving from electron spin 1 to 2.

The Hamiltonian given by Eq. (3) introduces the geometric phase shift under coherent optical excitation [20]. Thus if we start with a spin state $|s_0\rangle$ ($s_0 = \uparrow\uparrow, \uparrow\downarrow, \downarrow\uparrow$ or $\downarrow\downarrow$) of the electrons and the coherent state $|\alpha_0\rangle$ of the virtual polariton at $t = 0$, the state after a time $t$ is a coherent state with the same spin state $s_0$ but with a geometric phase factor i.e., $|\psi(t)\rangle = e^{i\phi(t)}|\alpha(t), s_0\rangle$. This is true even in the presence of polariton leakage from the cavity. However, if the initial state is a superposition of spin states of the localized electrons, then due to polariton leakage from the cavity, the system at a later time $t$ is a mixture of the states $\{|\alpha(t)_s, s\rangle\}$ and is best described by a density operator $\rho$. The master equation accounting for the finite lifetime of the polaritons at zero temperature is:

$$\frac{d\rho}{dt} = -i[H_0, \rho] + \gamma p \rho p^\dagger - \frac{\gamma}{2}\{p^\dagger p, \rho\}.$$

The decay of the coherence of the combined system due to polariton leakage is represented in the Lindblad form. After applying a Gaussian laser pulse with a full width at half maximum $\tau$ i.e., $\Omega(t) = \Omega e^{-(t^2/\tau^2)}$, the virtually excited polariton is in a time dependent coherent state $|\alpha(t)\rangle$. The time evolution of the polariton field is obtained by solving the master equation,

$$\alpha(t)_s = -i \int_0^t \Omega(s) e^{-i\delta_s(t-s)} e^{-\gamma(t-s)/2} ds \quad (4)$$

and the phase differene between the two spin states $|s_i\rangle$

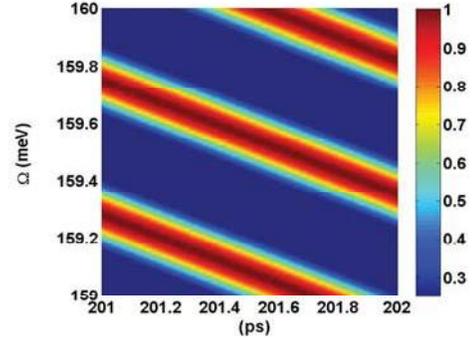

FIG. 2. Plot of fidelity in terms of the laser parameters $\tau$ and $\Omega$ when $\gamma = 3$ meV, $\delta_{\uparrow\uparrow} = 1$ meV, $\delta_{\uparrow\downarrow} = \delta_{\downarrow\uparrow} = 1.001$ meV and $\delta_{\downarrow\downarrow} = 1.002$ meV. The highest fidelity is 99.99% when $\tau = 201.88$ ps and $\Omega = 159.88$ meV.

and $|s_j\rangle$ is

$$\phi_{s_i,s_j}(t) = -\Re \int_0^t \Omega(t')\alpha_{s_i}(t')dt' + \Re \int_0^t \Omega(t')\alpha_{s_j}(t')dt'$$
$$- i\int_0^t \gamma \alpha_{s_i}(t')\alpha_{s_j}^*(t')dt'] + i\frac{\gamma}{2}\int_0^t |\alpha_{s_i}(t')|^2 dt'$$
$$+ i\frac{\gamma}{2}\int_0^t |\alpha_{s_j}(t')|^2 dt'. \quad (5)$$

The detailed derivation can be found in the supplementary information. It is clear from the last three terms of the above Eq. that a finite $\gamma$ introduces decoherence in the phase evolution.

If the initial state of the two electron spins is $\frac{1}{2}(|\uparrow\uparrow\rangle + |\downarrow\downarrow\rangle + |\uparrow\downarrow\rangle + |\downarrow\uparrow\rangle)$, we wish the final state is $\frac{1}{2}(|\uparrow\uparrow\rangle - |\downarrow\downarrow\rangle + |\uparrow\downarrow\rangle + |\downarrow\uparrow\rangle)$ which is a controlled $z$-operation between the electrons 1 and 2. The fidelity $F$ is given by:

$$F = Tr[\sqrt{\rho_a \rho_e \sqrt{\rho_a}}]$$
$$= \frac{1}{8}(3 - e^{-\Gamma_{s_1,s_2}}\cos(\Theta_{s_1,s_2}) + 2e^{-\Gamma_{s_1,s_3/s_4}}\cos(\Theta_{s_1,s_3/s_4})$$
$$- 2e^{-\Gamma_{s_2,s_3/s_4}}\cos(\Theta_{s_2,s_3/s_4})), \quad (6)$$

where, $\rho_a$ is the desired final density matrix, $\rho_e$ is the real density matrix obtained by considering errors introduced due to finite polariton lifetime, $\Theta_{s_i,s_j} = \Re[\phi_{s_i,s_j}]$ and $\Gamma_{s_i,s_j} = \Im[\phi_{s_i,s_j}]$. Here, $\Re$ and $\Im$ represent the real and imaginary parts.

Due to the decoherence caused by the finite lifetime of the polaritons, $F$ is generally reduced to below 1. However we can adjust the pump laser pulse duration $\tau$ and pulse energy, i.e., Rabi frequency $\Omega$ to get the best results. In Fig.2 the achievable fidelity is shown for the two laser parameters assuming the polariton lifetime $\approx$ 1.3 ps and $V = 2$ $\mu$eV. $F$ depends on $\phi_{s_i,s_j}$ and Eq.5 (with Eq.4) shows that to maintain the same phase difference, if $\Omega$ increases then $\tau$ must decrease. This is reflected in the stripe pattern in Fig.2. In addition, $F$ will be 1 when $\phi_{\downarrow\downarrow} - \phi_{\uparrow\downarrow(\downarrow\uparrow)} = (2n+1)\pi$ and $\phi_{\downarrow\downarrow} - \phi_{\uparrow\uparrow} = (2m+1)\pi$, which explains the alternating regions of high and low fidelity in Fig.2. The maximum fidelity of 99.99% is obtained for $\tau = 201.88$ ps and $\Omega = 159.88$ meV. Figure 3 shows the evolution of the coherent excitation induced by the pump pulse. The system returns to its original state within 4 times the pulse duration (thus, gate time $\tau_g = 4\tau$. The noise due to the vacuum field fluctuation is shown as a circle of radius 1/2 at the time when the three trajectories are most distinct from each other. We see that the trajectories lie well within quantum noise circles, preventing any loss of information to an eavesdropper. The pump power is given by, $P = \Omega(t)^2 \omega \tau_{photon}/\hbar$, which means when $V = 2$ $\mu$eV in order to achieve maximum $F$ the required peak power is $P$=9.7 mW. Thus with a reasonable laser power we can achieve a controlled-$z$ gate for the electrons 1 and 2.

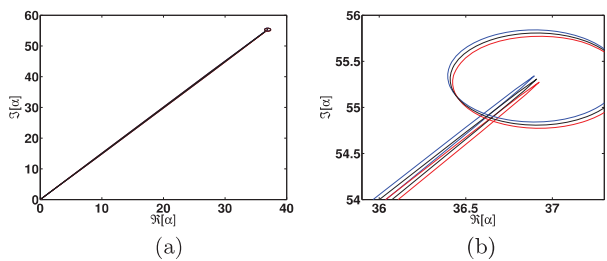

FIG. 3. (a) Plot of the evolution of the coherent state leaking from the cavity. The red curve represents the evolution when $|s\rangle = |\uparrow\uparrow\rangle$, the black curve when $|s\rangle = |\uparrow\downarrow\rangle$ (or $|\downarrow\uparrow\rangle$) and the blue curve when $|s\rangle = |\downarrow\downarrow\rangle$. The noise circles are plotted at the time when these curves are farthest from each other. (b) The magnified image of the noise circles.

The key advantage of the proposed scheme is the high speed of the two-qubit operation. Within 4 times the pulse duration (order of 800 ps) the desired phase rotations are achieved. This is about $10^6$ times faster than two-qubit gates in the proposed schemes for trapped ions [21, 22]. The moderate $Q$ values ($\approx 500$) required for the gate operation makes it a very practical setup, compared to the ones requiring much higher $Q$ values [23, 24]. By adjusting the laser parameters ($\delta_p$, $\Omega$ and $\tau$) we can achieve maximum fidelity for practical pump power.

In conclusion, we have outlined a two-qubit geometric phase gate, the performance of which can be controlled by the optical pulse shape. It promises a ultra-fast, robust and scalable all optical two qubit gate for electron spins.

### ACKNOWLEDGMENTS

This work has been supported by the Japan Society for the Promotion of Science(JSPS) through its Funding Program for World-Leading Innovative R&D on Science and Technology (FIRST Program)" and NICT. The authors appreciate Dr. G. F. Quinteiro, Dr. T. D. Ladd and Dr. Y. C. N. Na for insightful discussions.